\documentclass{webofc}
\usepackage[varg]{txfonts}
\usepackage{graphicx}
\usepackage[draft]{minted}
\usepackage{hyperref}

\begin{document}
\title{Awkward Arrays in Python, C++, and Numba}

\author{%
\firstname{Jim} \lastname{Pivarski}\inst{1}\fnsep\thanks{\email{pivarski@princeton.edu}} \and
\firstname{Peter} \lastname{Elmer}\inst{1}\fnsep\thanks{\email{Peter.Elmer@cern.ch}} \and
\firstname{David} \lastname{Lange}\inst{1}\fnsep\thanks{\email{David.Lange@cern.ch}}}

\institute{Princeton University}

\abstract{The Awkward Array library has been an important tool for physics analysis in Python since September 2018. However, some interface and implementation issues have been raised in Awkward Array's first year that argue for a reimplementation in C++ and Numba. We describe those issues, the new architecture, and present some examples of how the new interface will look to users. Of particular importance is the separation of kernel functions from data structure management, which allows a C++ implementation and a Numba implementation to share kernel functions, and the algorithm that transforms record-oriented data into columnar Awkward Arrays.}

\maketitle

\section{Introduction}

Columnar data structures, in which identically typed data fields are contiguous in memory, are a good fit to physics analysis use-cases. This was recognized as early as 1989 when column-wise ntuples were added to PAW and in 1997 when ``splitting'' was incorporated in the ROOT file format~\cite{rootio-1997}. In the past decade, with the Google Dremel paper~\cite{dremel}, the Parquet file format~\cite{parquet}, the Arrow memory interchange format~\cite{arrow}, and the inclusion of ``ragged tensors'' in TensorFlow~\cite{tf-raggedtensor}, the significance of hierarchical columnar data structures has been recognized beyond particle physics.

With the exception of the Columnar Objects experiment of T.\ Mattis et.\ al.~\cite{columnar-objects} and the XND library~\cite{xnd-io}, all of these projects focus on representing, storing, and transmitting columnar data structures, rather than operating on them. Physicists need to apply structure-changing transformations to search for decay topology candidates and other tasks that can change the level of nesting and multiplicity of their data. Operations of this complexity can be defined as a suite of primitives, allowing for NumPy-like convenience in Python~\cite{chep-2018}.

\begin{figure}
\begin{center}
\includegraphics[width=0.75\linewidth]{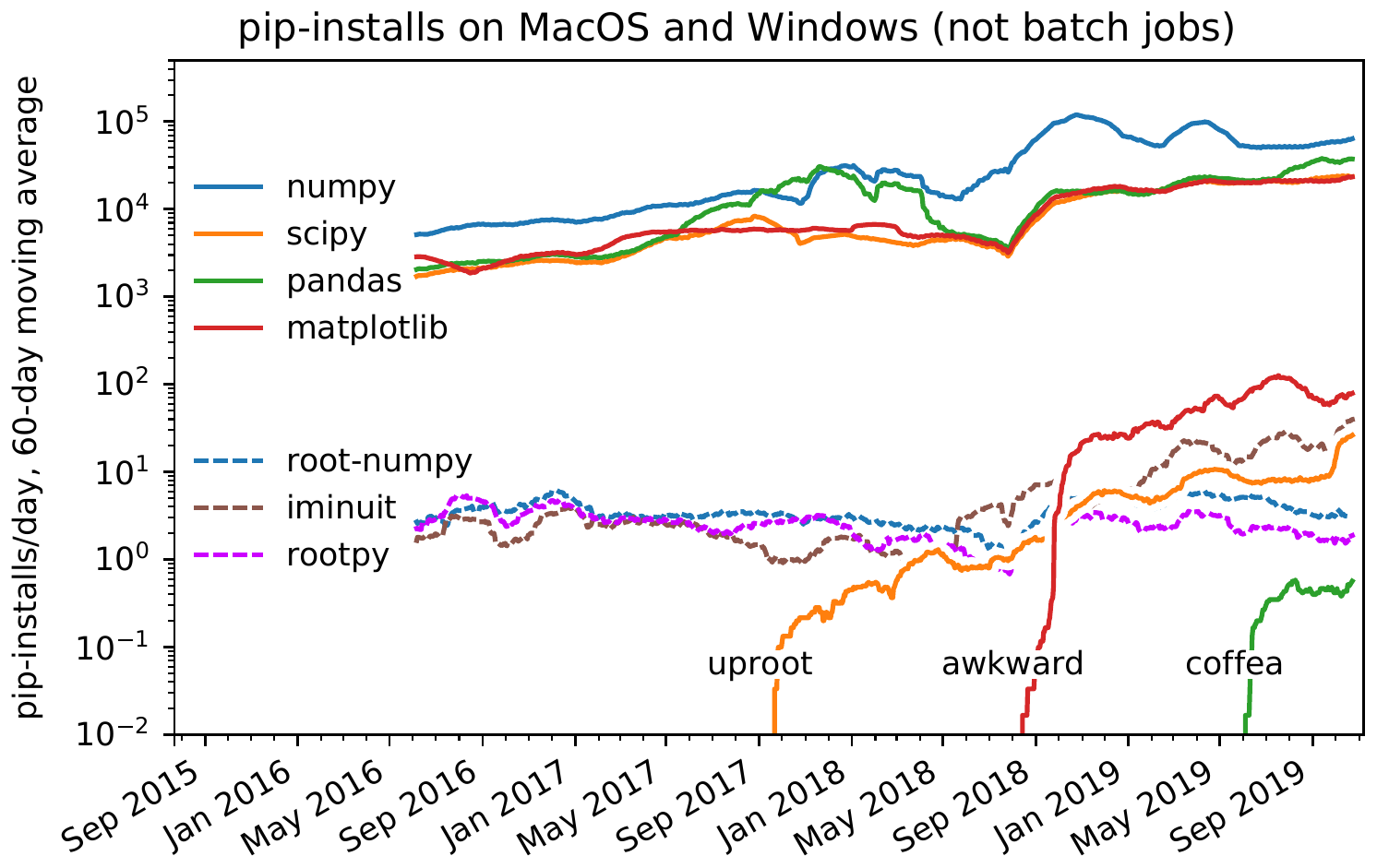}
\end{center}

\vspace{-0.25 cm}
\caption{Number of pip-installations per day (smoothed by a 60-day moving average) for popular data analysis libraries (numpy, scipy, pandas, matplotlib) and particle physics libraries (root-numpy, iminuit, rootpy, uproot, awkward, coffea) on operating systems not used for batch jobs (MacOS and Windows). \label{fig:pip-timeline}}
\end{figure}

The Awkward Array library~\cite{root-workshop-2018} was created to provide these operations on array objects that are easily convertible to the other libraries (zero-copy in some cases). Since its release in September 2018, Awkward Array has become one of the most widely pip-installed packages for particle physics (see Figure~\ref{fig:pip-timeline}).

Feedback from physicists, such as the interviews we reported previously~\cite{acat-2019} and in private conversations at a series of tutorials, has revealed that physicists appreciate the NumPy-like interface when it's easy to see how an analysis task can be expressed that way, but still need an interface for imperative programming. In addition, some names were poorly chosen, leading to confusion and name-conflicts, and more of the library's internal structure should be hidden from end-users. Also, the original library's pure NumPy implementation has been hard to extend and maintain.

All of these issues argue for a redesign of the library, keeping the core concepts that made it successful, restructuring the internals for maintainance, and presenting a simpler, more uniform interface to the user. This reimplementation project was dubbed ``Awkward 1.x'' and was completed in March 2020.

\section{Architecture}

The principle of Awkward Array is that an array of any data structure can be constructed from a composition of nodes that each provide one feature. The prototypical example is a jagged array, which represents an array of unequal-length subarrays with an array of integer \mintinline{c++}{offsets} and a contiguous array of \mintinline{c++}{content}. If the \mintinline{c++}{content} is one-dimensional, the jagged array is two-dimensional, where the second dimension has unequal lengths. To make a three-dimensional jagged array (unequal lengths in both inner dimensions), one jagged array node can be used as the \mintinline{c++}{content} for another. With an appropriate set of generators, any data structure can be assembled.

In the original Awkward Array library, the nodes were Python classes with special methods that NumPy recognizes to pass array-at-a-time operations through the data structure. Although that was an easy way to get started and respond rapidly to users' needs, some operations are difficult to implement in NumPy calls only. For complete generality, Awkward 1.x nodes are implemented as C++ classes, operated upon by specially compiled code.

We can satisfy the need for imperative access by adding Numba~\cite{numba} extensions to Awkward Array, but this would amount to rewriting the entire library, once in precompiled code (C++), and once in JIT-compiled code (Numba). To ease maintainance burdens, we have separated the code that implements operations from the code that manages data structures. Data structures are implemented twice---in C++ and Numba---but they both call the same suite of operations. In total, there are four layers:

\begin{enumerate}
\item High-level user interface in Python, which presents a single \mintinline{python}{awkward.Array} class.
\item Nested data structure nodes: C++ classes wrapped in Python with pybind11.
\item Two versions of the data structures, one in C++ and one in Numba.
\item Awkward Array operations in specialized, precompiled code with a pure C interface (can be called from C++ and Numba), called ``kernel functions.''
\end{enumerate}

\noindent With one exception to be discussed in Section~\ref{lab:fillablearray}, all loops that scale with the number of array elements are in the kernel functions layer. All allocation and memory ownership is in the C++ and Numba layer. This separation mimics NumPy itself, which uses Python reference counting to manage array ownership and precompiled code for all operations that scale with the size of the arrays. Also, like CuPy and array libraries for machine learning, adding GPU support would only require a new implementation of the kernel functions, not all layers.

\subsection{High-level Python layer}
\label{lab:high-level}

From a data analyst's perspective, the new Awkward Array library has only one important data type, \mintinline{python}{awkward.Array}, and a suite of functions operating on that type.

\begin{minted}{python}
>>> import awkward as ak
>>> array = ak.Array([[{"x": 1, "y": [1.1]}, {"x": 2, "y": [2.0, 0.2]}],
...                   [], [{"x": 3, "y": [3.0, 0.3, 3.3]}]])
>>> array
<Array [[{x: 1, y: [1.1]}, ... 3.3]}]] type='3 * var * {"x": int64,...'>
\end{minted}

\noindent Much like NumPy's \mintinline{python}{dtype}, the actual type of the array is a Python value (presented in DataShape~\cite{datashape} notation).

\begin{minted}{python}
>>> ak.typeof(array)
3 * var * {"x": int64, "y": var * float64}
\end{minted}

\noindent These arrays can be sliced like NumPy arrays, with a mix of integers, slices, arrays of booleans and integers, jagged arrays of booleans and integers, but for any data structure.

\begin{minted}{python}
>>> array["y", [0, 2], :, 1:]
<Array [[[], [0.2]], [[0.3, 3.3]]] type='2 * var * var * float64'>
\end{minted}

\noindent They can also be operated on with NumPy's array-at-a-time functions.

\begin{minted}{python}
>>> import numpy as np
>>> np.sin(array)
<Array [[{x: 0.841, ... -0.158]}]] type='3 * var * {"x": float64,...'>
\end{minted}

\begin{figure}
\begin{center}
\includegraphics[width=\linewidth]{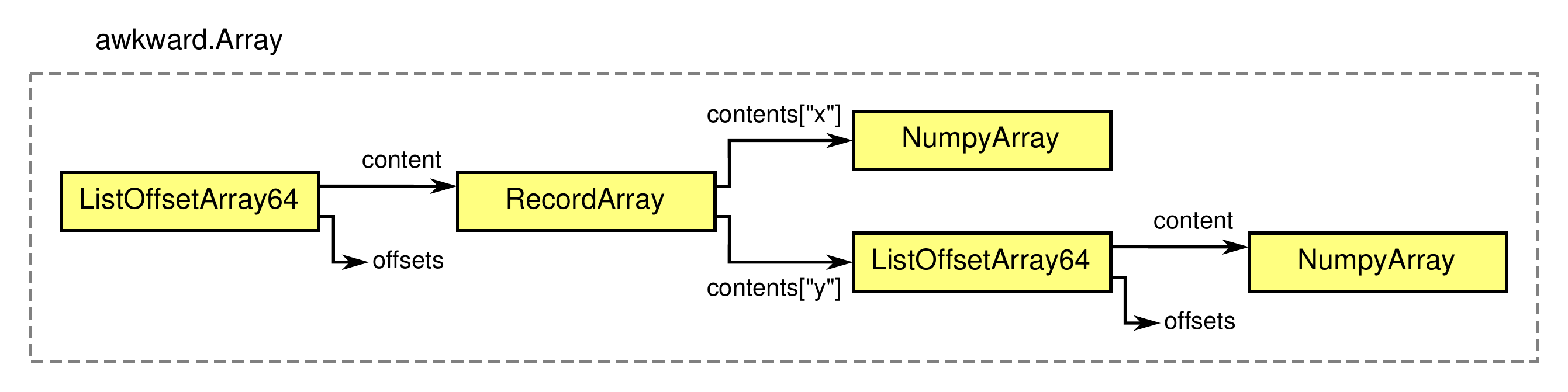}
\end{center}

\vspace{-0.25 cm}
\caption{Structure of the array discussed in Section~\ref{lab:high-level}: hierarchical \mintinline{python}{layout} nodes are wrapped in a single, user-facing \mintinline{python}{awkward.Array}. \label{fig:example-hierarchy}}
\end{figure}

The nodes that define the structure of an array, which were user-level types in the original Awkward Array library, are accessible through a \mintinline{python}{layout} property, illustrated in Figure~\ref{fig:example-hierarchy}. Most physicists won't need to use these nodes directly.

A surprisingly important use-case for the original Awkward Array was the ability to add domain-specific code to the data structures. For example, records with fields named \mintinline{python}{"pt"}, \mintinline{python}{"eta"}, \mintinline{python}{"phi"}, and \mintinline{python}{"mass"} were wrapped as \mintinline{python}{LorentzVector} objects, with operations on arrays of \mintinline{python}{LorentzVectors} (addition, boosting, $\Delta R$ distances, etc.) provided as methods. However, this feature was implemented using Python class inheritance, which was fragile because new Python objects for the same data are frequently created, and it was easy to lose the necessary superclasses in these transformations.

This feature is implemented in Awkward 1.x by instead applying the interpretation only when creating the high-level wrapper, and keeping track of how to interpret each \mintinline{python}{layout} node with JSON-formatted \mintinline{python}{parameters} that pass through C++ and Numba. For example,

\begin{minted}{python}
>>> class Point(ak.Record):
...   def __repr__(self):
...     return "Point({} {})".format(self["x"], self["y"])
\end{minted}

\begin{minted}{python}
>>> ak.namespace["Point"] = Point
>>> array.layout.content.setparameter("__class__", "Point")
>>> array.layout.content.setparameter("__str__", "P")
>>> array
<Array [[Point(1 [1.1]), ... Point(3 [3, 0.3, 3.3])]] type='3 * var * P'>
\end{minted}

\noindent Incidentally, strings are implemented the same way: there is no string array type, but lists of 8-bit integers that should be interpreted as strings are labeled as such, and therefore presented and operated upon as such.

\begin{minted}{python}
>>> array = ak.Array(["one", "two", "three"])
>>> ak.tolist(array)      # with the string interpretation
['one', 'two', 'three']
\end{minted}

\begin{minted}{python}
>>> array.layout.content.setparameter("__class__", None)
>>> ak.tolist(array)      # without the string interpretation
[[111, 110, 101], [116, 119, 111], [116, 104, 114, 101, 101]]
\end{minted}

\subsection{C++ layer}

The \mintinline{python}{layout} nodes underlying the above example are all C++ class instances, wrapped in Python using pybind11~\cite{pybind11} (see Figure~\ref{fig:example-hierarchy}). To allow dynamic array construction in Python, these nodes are reference-counted with virtual inheritance: \mintinline{c++}{std::shared_ptr<Content>}, where \mintinline{c++}{Content} is the superclass of all node classes. Compile-time templates are only used to specialize integer types (e.g.\ \mintinline{c++}{ListOffsetArray32} versus \mintinline{c++}{ListOffsetArray64}), not for building nested structures.

The use of shared pointers and virtual inheritance might, at first, seem to be a performance bottleneck, but it is not. An operation on an Awkward Array only needs to step through the shared pointers and inheritance that defines the data {\it type}, which is several to hundreds of nodes at most. The same operation loops over the {\it values} in the array, which can number in the billions, in the kernel functions, which involve no smart pointers or inheritance. Thus, optimization efforts should focus on the kernel functions, rather than the C++ layer.

\subsection{Numba layer}

Numba is an opt-in JIT-compiler for a subset of Python, extensively covering NumPy arrays and their operations. Since Numba-compiled code looks like familiar, imperative Python, users can debug algorithms without compilation and only JIT-compile those functions when they are ready to scale up to large datasets.

Numba has an extension mechanism that allows third-party libraries to inform the Numba compiler of new data types. Awkward 1.x uses this extension mechanism to implement Awkward Arrays and their operations in Numba-compiled functions. This is a second implementation of the node data structures and their memory-ownership, but not the kernel functions, which C++ and Numba both call.

\subsection{Kernel functions layer}

All operations that transform arrays are implemented in a suite of kernel functions, which are written in C++ but exported as \mintinline{c++}{extern "C"}. Only C-language features can be used in the function signatures, which excludes classes, dispatch by argument types, and templates. Although this is inconvenient, Numba can only call external C functions, not C++, and thus this is a requirement for C++ and Numba to use the same kernel functions.

Apart from internal template specialization on some argument types, the kernel function implementations also resemble pure C functions because they consist entirely of \mintinline{c++}{for} loops that fill preallocated arrays (allocated and owned by C++ or Numba). Our use of the word ``kernel'' derives from the fact that this separation between slow bookkeeping in C++ and fast math in simple, C-like code resembles the separation of CPU-bound and GPU-bound code in GPU applications. Thus, the library is already organized in a GPU-friendly way; all that remains is to provide GPU-native implementations of each kernel function.

All foreseeable optimization effort will be focused on the kernel functions, rather than the bookkeeping and interface code in C++, Numba, and Python, with one exception: record-oriented $\to$ columnar data transformations discussed in the next section.

\section{Record-oriented $\to$ columnar}
\label{lab:fillablearray}

Transformations of Awkward Arrays to and from columnar formats like Arrow and ``split'' ROOT branches are either single-\mintinline{c++}{malloc} array copies or zero-copy data casting. Record-oriented data, however, require significant processing to transform into any columnar format.

\begin{figure}
\begin{center}
\includegraphics[width=0.75\linewidth]{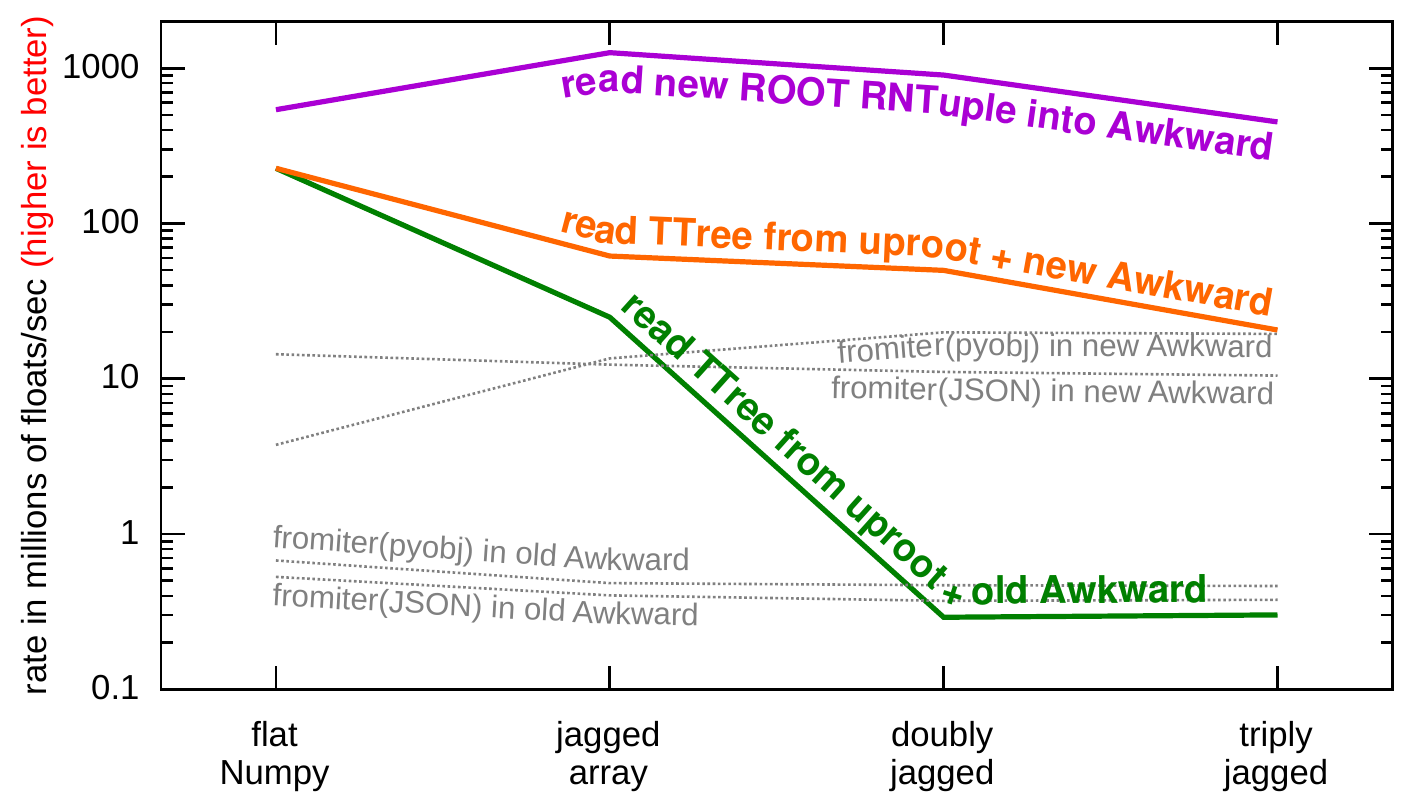}
\end{center}

\vspace{-0.25 cm}
\caption{Rate of reading {\tt list}$^{\mbox{\tt N}}${\tt (float)} data from Python objects (``pyobj'') and from JSON strings in the old and new Awkward \mintinline{python}{fromiter}, from ROOT's old \mintinline{c++}{TTree}, which only needs the transformation for doubly jagged and above, and from ROOT's new \mintinline{c++}{RNTuple}, which never needs the transformation. \label{fig:read_ttree}}
\end{figure}

Such a function, named \mintinline{python}{fromiter} in the original Awkward Array library, had many important use-cases. We have therefore moved the \mintinline{python}{fromiter} implementation from Python into C++ and observe a 10--20$\times$ speed-up for typical data structures (see Figure~\ref{fig:read_ttree}). Any record-oriented $\to$ columnar transformation of data whose type is not known at compile-time must include virtual method indirection, so further optimization is only possible for specialized types. This is the exception to the rule that all operations that scale with the size of the dataset must be implemented in kernel functions, because the accumulated arrays are dynamically typed.

Our record-oriented $\to$ columnar algorithm discovers the data's type during the data transformation pass. For example, if a particular field has always been filled with integers, the first time it is filled with a floating-point value invokes a conversion of the previous integer data into floating-point, then the floating-point array is used henceforth. In the same example, later filling that field with a string invokes a replacement of the floating-point array with a tagged union of floating-point and strings (reusing the floating-point array).

Awkward 1.x provides three interfaces to this algorithm: (a)~Python objects $\to$ Awkward Arrays, like the old \mintinline{python}{fromiter}, (b)~JSON data $\to$ Awkward Arrays using the RapidJSON C++ library's SAX interface, and (c)~a builder pattern in which the user can fill individual values via method calls. The latter is the most powerful interface, and it is provided in Python, C++, and Numba. In C++, it would allow Awkward interfaces for established C++ projects, and in Numba, it provides a convenient way for physicists to build complex data structures.

One particularly important special case of record-oriented $\to$ columnar transformation is unequal-length lists (of any depth of nesting) of numbers. Many analysis-level ROOT files contain data of this type, though ROOT's \mintinline{c++}{TTree} serialization only stores singly jagged arrays in a columnar format: deeper levels are record-oriented. Since the data type is partially known, a specialized implementation improves upon both the old and new \mintinline{python}{fromiter}, as shown in Figure~\ref{fig:read_ttree}. The more long-term solution, however, is ROOT's new \mintinline{c++}{RNTuple} serialization, which is columnar at all levels, making this transformation unnecessary.


\section{Transitioning to the new library}

The reimplementation of Awkward Array in C++ and Numba provides new features, a more unified interface, and higher performance in some cases. These improvements are being introduced to users as a new library to ease the transition. The old library can still be pip-installed/imported as \mintinline{bash}{awkward}, whereas the new one is available as \mintinline{bash}{awkward1}. Uproot is transitioning in a similar way, with a new \mintinline{bash}{uproot4} using \mintinline{bash}{awkward1} and the original \mintinline{bash}{uproot} using \mintinline{bash}{awkward}.

Once adoption of the new libraries increases, they will become defaults by renaming \mintinline{bash}{awkward1}/\mintinline{bash}{uproot4} as \mintinline{bash}{awkward}/\mintinline{bash}{uproot}, with the originals becoming \mintinline{bash}{awkward0} and \mintinline{bash}{uproot3}. Thus, legacy scripts (perhaps necessary for a student's graudation) can be kept functional by adding

\begin{minted}{python}
import awkward0 as awkward
import uproot3 as uproot
\end{minted}

\noindent while new scripts default to the new libraries. We expect this transition to be complete by the end of 2020.

\section{Acknowledgements}

Support for this work was provided by NSF cooperative agreement OAC-1836650 (IRIS-HEP), grant OAC-1450377 (DIANA/HEP) and PHY-1520942 (US-CMS LHC Ops).

\end{document}